\documentclass{article}
\usepackage{amsmath, amssymb, amsfonts}
\title{Comment on ''Empty Black Holes, Firewalls, and the Origin of Bekenstein-Hawking Entropy''} 
\author{Hristu Culetu, \\Ovidius University, Dept.of Physics and Electronics, \\B-dul Mamaia 124, 900527 Constanta, Romania, \\e-mail : hculetu@yahoo.com}

\begin{document}
\numberwithin{equation}{section}
\pagenumbering{arabic}
\maketitle
\newcommand{\fv}{\boldsymbol{f}}
\newcommand{\tv}{\boldsymbol{t}}
\newcommand{\gv}{\boldsymbol{g}}
\newcommand{\OV}{\boldsymbol{O}}
\newcommand{\wv}{\boldsymbol{w}}
\newcommand{\WV}{\boldsymbol{W}}
\newcommand{\NV}{\boldsymbol{N}}
\newcommand{\hv}{\boldsymbol{h}}
\newcommand{\yv}{\boldsymbol{y}}
\newcommand{\RE}{\textrm{Re}}
\newcommand{\IM}{\textrm{Im}}
\newcommand{\rot}{\textrm{rot}}
\newcommand{\dv}{\boldsymbol{d}}
\newcommand{\grad}{\textrm{grad}}
\newcommand{\Tr}{\textrm{Tr}}
\newcommand{\ua}{\uparrow}
\newcommand{\da}{\downarrow}
\newcommand{\ct}{\textrm{const}}
\newcommand{\xv}{\boldsymbol{x}}
\newcommand{\mv}{\boldsymbol{m}}
\newcommand{\rv}{\boldsymbol{r}}
\newcommand{\kv}{\boldsymbol{k}}
\newcommand{\VE}{\boldsymbol{V}}
\newcommand{\sv}{\boldsymbol{s}}
\newcommand{\RV}{\boldsymbol{R}}
\newcommand{\pv}{\boldsymbol{p}}
\newcommand{\PV}{\boldsymbol{P}}
\newcommand{\EV}{\boldsymbol{E}}
\newcommand{\DV}{\boldsymbol{D}}
\newcommand{\BV}{\boldsymbol{B}}
\newcommand{\HV}{\boldsymbol{H}}
\newcommand{\MV}{\boldsymbol{M}}
\newcommand{\be}{\begin{equation}}
\newcommand{\ee}{\end{equation}}
\newcommand{\ba}{\begin{eqnarray}}
\newcommand{\ea}{\end{eqnarray}}
\newcommand{\bq}{\begin{eqnarray*}}
\newcommand{\eq}{\end{eqnarray*}}
\newcommand{\pa}{\partial}
\newcommand{\f}{\frac}
\newcommand{\FV}{\boldsymbol{F}}
\newcommand{\ve}{\boldsymbol{v}}
\newcommand{\AV}{\boldsymbol{A}}
\newcommand{\jv}{\boldsymbol{j}}
\newcommand{\LV}{\boldsymbol{L}}
\newcommand{\SV}{\boldsymbol{S}}
\newcommand{\av}{\boldsymbol{a}}
\newcommand{\qv}{\boldsymbol{q}}
\newcommand{\QV}{\boldsymbol{Q}}
\newcommand{\ev}{\boldsymbol{e}}
\newcommand{\uv}{\boldsymbol{u}}
\newcommand{\KV}{\boldsymbol{K}}
\newcommand{\ro}{\boldsymbol{\rho}}
\newcommand{\si}{\boldsymbol{\sigma}}
\newcommand{\thv}{\boldsymbol{\theta}}
\newcommand{\bv}{\boldsymbol{b}}
\newcommand{\JV}{\boldsymbol{J}}
\newcommand{\nv}{\boldsymbol{n}}
\newcommand{\lv}{\boldsymbol{l}}
\newcommand{\om}{\boldsymbol{\omega}}
\newcommand{\Om}{\boldsymbol{\Omega}}
\newcommand{\Piv}{\boldsymbol{\Pi}}
\newcommand{\UV}{\boldsymbol{U}}
\newcommand{\iv}{\boldsymbol{i}}
\newcommand{\nuv}{\boldsymbol{\nu}}
\newcommand{\muv}{\boldsymbol{\mu}}
\newcommand{\lm}{\boldsymbol{\lambda}}
\newcommand{\Lm}{\boldsymbol{\Lambda}}
\newcommand{\opsi}{\overline{\psi}}
\renewcommand{\tan}{\textrm{tg}}
\renewcommand{\cot}{\textrm{ctg}}
\renewcommand{\sinh}{\textrm{sh}}
\renewcommand{\cosh}{\textrm{ch}}
\renewcommand{\tanh}{\textrm{th}}
\renewcommand{\coth}{\textrm{cth}}

\begin{abstract}
Saravani, Afshordi and Mann \cite{SAM} considered a surface fluid with vanishing energy density on the stretched horizon of a black hole, taken as the new boundary of spacetime. We show that their entropy per unit area of the fluid does not correspond to the Bekenstein-Hawking entropy, both for a Schwarzschild black hole and for a Kerr-Newman black hole. Due to the lack of a horizon, we argue that the Unruh effect could not be taken into account and the concept of local thermodynamic equilibrium is not sustained because of the $\theta$-dependence of the temperature.\\
\end{abstract}

 In the paper \cite{SAM}, Saravani, Afshordi and Mann (SAM) proposed a model for black holes (BHs) where the spacetime ends at a microscopic distance from the BH horizon. They put a (2+1) dimensional surface fluid with vanishing energy density at the boundary (stretched horizon) and obtained a thermodynamic entropy for the fluid identical to the Bekenstein-Hawking entropy. 
 
 Let us observe that, even though the Sec. II.A is called "Schwarzschild Black Holes", their geometry (2)
 \begin{equation}
 ds^{2} =- N^{2}(r)dt^{2} + \frac{dr^{2}}{f(r)} + r^{2} d \Omega^{2},
\label{1}
\end{equation}
has no any horizon (SAM have introduced a boundary at $r = r^{*} > r_{0}$, where $r_{0}$ stands for the horizon, with $N(r_{0}) = 0)$. The Authors of \cite{SAM} keep the above metric in its general form throughout the Sec. II.A. Not possessing a horizon, their metric (2), in our opinion, cannot generate thermal radiation (Unruh's effect) even though a static observer on the surface fluid is accelerating (the spacetime is curved and has spherical symmetry). We could eventually sit close to the horizon of the BH where the metric is of Rindler-type. SAM should have used the same approximation as in Sec. B (for Kerr-Newman BHs), going near the horizon where the spacetime is Minkowskian and the appearance of the Unruh radiation is natural. Therefore, saying that ''Since the surface fluid is at constant radius it consequently sees the thermal radiation due to its acceleration (Unruh effect)'' \cite{SAM} (below Eq. (14)) is, in our view, incorrect as long as the authors' spacetime terminates at a distance $r^{*} > r_{0}$, namely the horizon is left outside. In contrast, at Sec. II.B, SAM looked from the very beginning for the near horizon approximation for the Kerr-Newman geometry (their 1st equation from Sec. B), where $\lambda = 0$ corresponds to the BH horizon. From Eq. (B1) it is clear that $\lambda$ (as the time $\tau$) is dimensionless. Therefore, it seems to be a typos in the integral expression of it and, to respect the units, we must have
  \begin{equation}
  \lambda = \int^{r}_{r_{+}} \frac{dr'}{\sqrt{\Delta(r')}}
\label{2}
\end{equation}
Anyway, only with this form of $\lambda$ the result $\lambda \approx 2\sqrt{(r - r_{+})(r_{+} - r_{-})} + O((r - r_{+})^{3/2})$ holds.

As far as Eqs. (13) and (14) are concerned, they have been previously written by Mazur and Mottola \cite{MM} and Ghezzi \cite{CG} using the same Israel junction conditions ($\eta$ from \cite{MM} and \cite{CG} correspond to $\Sigma$ from \cite{SAM}, $\Pi$ from \cite{SAM} is $\sigma$ from \cite{MM} or $-\sigma$ from \cite{CG} and $N^{2}$ and $f$ from \cite{SAM} correspond to $f$ and, respectively $h$ from \cite{MM}). We notice also that, with $r^{*} \neq r_{0}$, (17) cannot be obtained from (16), using (13) and (14). One obtains in fact
  \begin{equation}
  S = \frac{1}{4} - \frac{N(r^{*})}{4r^{*}N'(r^{*})}
\label{3}
\end{equation}
where $N'(r) = dN/dr$. Only when $r^{*} \rightarrow r_{0},~S = 1/4$ emerges. But that is not specified in \cite{SAM} and, anyway, it is trivial because $r_{0}$ corresponds to the event horizon (located at $r < r^{*}$), not to the stretched horizon. Nowhere SAM used that $r^{*} - r_{0} << r_{0}$, so their equations 13 - 16 are valid for any $r^{*} > r_{0}$. Let us ckeck the relation (0.3) for the Schwarzschild metric, namely $N^{2} = f = 1- (2m/r)$. We have $N/4rN' = f/2rf'$, so that 
  \begin{equation}
  s(r) = \frac{1}{4} - \frac{r}{4m}(1 - \frac{2m}{r}).
\label{4}
\end{equation}
We obtain, of course, that $s = 1/4$ at $r = r_{0} = 2m$ but $s$ vanishes for $r = 3m$ and, what is worse, it becomes negative for $r > 3m$. Although SAM stated that $r^{*}$ is at a microscopic distance from $r_{0}$, that constraint does not enter in any way in the calculations of $\Sigma, \Pi, T$ or $s$ from the Sec. II.A.

Due to the near horizon approximation used, SAM obtained directly (from their Lanczos equation (12)) the fact that $\Sigma = 0$ on the stretched horizon $\lambda = \lambda^{*}$ (Eq. (25)) and $\Pi \neq 0$ (note that in (13), $\Sigma = 0$ only for the Schwarzschild BH). Taking the limit $\lambda^{*} \rightarrow 0$ (the Kerr-Newman horizon), SAM correctly obtained $\Pi \rightarrow \infty$, as in \cite{KP} (Eq. 40), \cite{KKP} (Eq. 12) and \cite{HC} (Eq. 4.9), for a Schwarzschild-type spacetime, as the metric (2) from \cite{SAM} is. We stress again that, in our opinion, there is no any Unruh radiation here because the near horizon approximation (B1) of the Kerr-Newman BH is not flat (the Riemann tensor is nonzero, contrary to the Schwarzschild case where the metric is Rindler, close to its horizon). In addition, their formula (26) does not seem to be valid. Because of the fact that $\Gamma_{+} = r_{+}^{2} + \bar{a}^{2}cos^{2}\theta$ is $\theta$-dependent (where $r_{+} = m + \sqrt{m^{2} - \bar{a}^{2} - q^{2}}$), apart from the $\lambda$-component of the fluid acceleration $a^{\lambda} = 1/\Gamma_{+} \lambda$, there is also a nonzero $\theta$-componenent, $a^{\theta} = -(1/\Gamma_{+}^{2}) \bar{a}^{2} sin\theta cos\theta$, where $\bar{a} = J/m$. Therefore,
   \begin{equation}
   a \equiv \sqrt{a^{b} a_{b}} = \sqrt{\frac{1}{\Gamma_{+} \lambda^{2}} + \frac{\bar{a}^{4} sin^{2} \theta cos^{2} \theta}{\Gamma_{+}^{3}}}
\label{5}
\end{equation}
(Note that the Authors of \cite{SAM} use the same letter $a$ for the acceleration of the fluid and for $J/m$). Their expression (26) should be recovered only if $a^{\theta}$ were zero. As a consequence, the entropy (28) per unit area of the fluid is not equal to $1/4$ as they claim. Therefore, author's conjecture (at the beginning of Sec. IV) that ''the microstates of a black hole are those of the surface fluid at the stretched horizon'' should not be valid. From our point of view, the microstates of a BH are to be sought on its event horizon even if we agree with SAM that ''the membrane properties are physical and result from condensation of accreted matter into a physical membrane'' (Sec. IV). Moreover, authors' conclusion from the beginning of Sec. IV that the energy density $\Sigma$ is vanishing on the stretched horizon seems to be in contradiction with their Eq. (13) where $\Sigma \neq 0$ at $r = r^{*}$ (it vanishes only when $r^{*}  \rightarrow r_{0}$, as SAM themselves found below Eq. (14)). 

 The introduction of ''local thermal equilibrium'' is questionable since the spatial/temporal variation of $T(\textbf{x},t)$ should be small and this is not obeyed, for instance, by Eq. (27), where there are restrictions for $\lambda^{*}$ to be close to $\lambda = 0$ but not for angular variable $\theta$. In other words, we consider that, when $\theta$ varies, for example, from 0 to $\pi/2$, the variation of $T$ from (27) is not small (the ''local approximation'' is not valid when $\theta$ is varied on its entire range), so that local equilibrium at Unruh temperature $T = a/2 \pi$ cannot be set by an angle-dependent acceleration.

We would like to add a remark on the title of \cite{SAM}. If Authors' BHs are empty, why to call them ''Schwarzschild Black Holes''?. If the spacetime ends at the stretched horizon $r = r^{*}$, what is the source of curvature in the bulk? The surface stress tensor (3) refers to, as SAM tell us, to a thin layer of fluid sitting at the boundary. They do not give any informations about the bulk stress tensor but only that $N(r)$ and $f(r)$ are arbitrary functions satisfying GR field equations in the bulk (below Eq. (2)).


\begin{thebibliography} {6} 

\bibitem{SAM}
M. Saravani, N. Afshordi and R. Mann, arXiv: 1212.4176 [hep-th].
\bibitem{MM}
P. O. Mazur and E. Mottola, Proc. Nat. Acad. Sci.101, 9545 (2004) (arXiv: gr-qc/0407075).  
\bibitem{CG}
C. R. Ghezzi, Astrophys. Space Sci.333, 437 (2011) (arXiv: 0908.0779 [gr-qc]. 
\bibitem{KP}
S. Kolekar and T. Padmanabhan, Phys. Rev.D85, 024004 (2011) (arXiv: 1109.5353 [gr-qc].
\bibitem{KKP}
S. Kolekar, D. Kothawala and T. Padmanabhan, Phys. Rev.D85, 064031 (2012) (arXiv: 1111.0973 [gr-qc].
\bibitem{HC}
H. Culetu, Phys. Lett. A 376, 2817 (2012), (arXiv: 1103.2645 [gr-qc]).


\end{thebibliography}
\end{document}